\documentclass[twocolumn,times]{aastex6}
\usepackage{graphicx}
\usepackage{bm}
\usepackage{epsfig}
\usepackage{multirow}
\usepackage{color}
\usepackage{subfigure}

\def\hideprivate{\global\long\def\private##1{\iffalse ##1 \fi}}
\hideprivate

\defcitealias{GCN18330}{18330}
\defcitealias{GCN18331}{18331}
\defcitealias{GCN18332}{18332}
\defcitealias{GCN18333}{18333}
\defcitealias{GCN18334}{18334}
\defcitealias{GCN18335}{18335}
\defcitealias{GCN18336}{18336}
\defcitealias{GCN18337}{18337}
\defcitealias{GCN18338}{18338}
\defcitealias{GCN18339}{18339}
\defcitealias{GCN18340}{18340}
\defcitealias{GCN18341}{18341}
\defcitealias{GCN18343}{18343}
\defcitealias{GCN18344}{18344}
\defcitealias{GCN18345}{18345}
\defcitealias{GCN18346}{18346}
\defcitealias{GCN18347}{18347}
\defcitealias{GCN18348}{18348}
\defcitealias{GCN18349}{18349}
\defcitealias{GCN18350}{18350}
\defcitealias{GCN18353}{18353}
\defcitealias{GCN18354}{18354}
\defcitealias{GCN18359}{18359}
\defcitealias{GCN18361}{18361}
\defcitealias{GCN18362}{18362}
\defcitealias{GCN18363}{18363}
\defcitealias{GCN18364}{18364}
\defcitealias{GCN18370}{18370}
\defcitealias{GCN18371}{18371}
\defcitealias{GCN18372}{18372}
\defcitealias{GCN18388}{18388}
\defcitealias{GCN18390}{18390}
\defcitealias{GCN18394}{18394}
\defcitealias{GCN18395}{18395}
\defcitealias{GCN18397}{18397}
\defcitealias{GCN18420}{18420}
\defcitealias{GCN18424}{18424}
\defcitealias{GCN18474}{18474}
\defcitealias{GCN18655}{18655}
\defcitealias{GCN18690}{18690}
\defcitealias{GCN18709}{18709}
\defcitealias{GCN18851}{18851}
\defcitealias{GCN18858}{18858}
\defcitealias{GCN18903}{18903}
\defcitealias{GCN18914}{18914}
\defcitealias{GCN19013}{19013}
\defcitealias{GCN19017}{19017}
\defcitealias{GCN19021}{19021}
\defcitealias{GCN19022}{19022}
\defcitealias{GCN19034}{19034}
\newcommand\citegcn[1]{\citetalias{GCN#1}}

\begin{document}

\title{\lowercase{i}PTF Search for an Optical Counterpart to Gravitational Wave Trigger GW150914}
\author{M. M. Kasliwal \altaffilmark{1}, 
S. B. Cenko  \altaffilmark{2, 3}, 
L. P. Singer  \altaffilmark{2, 4},
A. Corsi  \altaffilmark{5},
Y. Cao \altaffilmark{1},
T. Barlow \altaffilmark{1}, 
V. Bhalerao \altaffilmark{6}, 
E. Bellm \altaffilmark{1}, 
D. Cook \altaffilmark{1},
G. E. Duggan \altaffilmark{1},
R. Ferretti  \altaffilmark{7},
D. A. Frail \altaffilmark{8},
A. Horesh \altaffilmark{9},
R. Kendrick \altaffilmark{10},
S. R. Kulkarni \altaffilmark{1},
R. Lunnan \altaffilmark{1},
N. Palliyaguru  \altaffilmark{5}, 
R. Laher \altaffilmark{11},
F. Masci \altaffilmark{12},
I. Manulis \altaffilmark{9},
A. A. Miller \altaffilmark{1,13,14}, 
P. E. Nugent \altaffilmark{15,16},
D. Perley \altaffilmark{17},
T. A. Prince \altaffilmark{1},
R. M. Quimby \altaffilmark{18,19},
J. Rana \altaffilmark{6}, 
U. Rebbapragada \altaffilmark{13},
B. Sesar \altaffilmark{20},
A. Singhal \altaffilmark{6}, 
J. Surace \altaffilmark{11},
A. Van Sistine \altaffilmark{21}
}

\altaffiltext{1}{Cahill Center for Astrophysics, California Institute of Technology, Pasadena, CA 91125, USA}
\altaffiltext{2}{Astrophysics Science Division, NASA Goddard Space Flight Center, Code 661, Greenbelt, MD 20771, USA}
\altaffiltext{3}{Joint Space-Science Institute, University of Maryland, College Park, MD 20742, USA}
\altaffiltext{4}{NASA Postdoctoral Program Fellow}
\altaffiltext{5}{Texas Tech University, Physics Department, Lubbock, TX 79409-1051, USA}
\altaffiltext{6}{Inter-University Centre for Astronomy and Astrophysics (IUCAA), Post Bag 4, Ganeshkhind, Pune 411007, India}
\altaffiltext{7}{The Oskar Klein Centre, Department of Physics, Stockholm University, SE-106 91 Stockholm, Sweden}
\altaffiltext{8}{National Radio Astronomy Observatory, Socorro NM}
\altaffiltext{9}{Department of Particle Physics and Astrophysics, Weizmann Institute of Science, 76100 Rehovot, Israel}
\altaffiltext{10}{Lockheed Martin Space Systems Company, Palo Alto,  CA}
\altaffiltext{11}{Spitzer Science Center, California Institute of Technology,  M/S 314-6, Pasadena, CA 91125, U.S.A.}
\altaffiltext{12}{Infrared Processing and Analysis Center, California Institute of Technology, Pasadena, CA 91125, USA}
\altaffiltext{13}{Jet Propulsion Laboratory, California Institute of Technology, Pasadena, CA 91109, USA}
\altaffiltext{14}{Hubble Fellow}
\altaffiltext{15}{Astronomy Department, University of California at Berkeley, Berkeley, CA 94720, USA}
\altaffiltext{16}{Lawrence Berkeley National Laboratory, 1 Cyclotron Road MS 50B-4206, Berkeley, CA 94720, USA}
\altaffiltext{17}{Dark Cosmology Centre, Niels Bohr Institute, Juliane Maries Vej 30, Copenhagen ¯, DK-2100, Denmark}
\altaffiltext{18}{San Diego State University, San Diego CA}
\altaffiltext{19}{Kavli IPMU (WPI), UTIAS, The University of Tokyo, Kashiwa, Chiba 277-8583, Japan}
\altaffiltext{20}{Max Planck Institute for Astronomy, K\"{o}nigstuhl 17, D-69117 Heidelberg, Germany}
\altaffiltext{21}{Department of Physics, University of Wisconsin-Milwaukee, Milwaukee, WI 53201, USA}

\begin{abstract}
The intermediate Palomar Transient Factory (iPTF) autonomously responded to and promptly tiled the error region of the first gravitational wave event GW150914 to search for an optical counterpart. Only a small fraction of the total localized region was immediately visible in the Northern night sky, due both to sun-angle and elevation constraints. Here, we report on the transient candidates identified and rapid follow-up undertaken to determine the nature of each candidate. Even in the small area imaged of 126 deg$^{2}$, after extensive filtering, 8 candidates were deemed worthy of additional follow-up. Within two hours, all 8 were spectroscopically classified by the Keck II telescope. Curiously, even though such events are rare, one of our candidates was a superluminous supernova. We obtained radio data with the Jansky Very Large Array and X-ray follow-up with the \textit{Swift} satellite for this transient. None of our candidates appear to be associated with the gravitational wave trigger, which is unsurprising given that GW150914 came from the merger of two stellar-mass black holes. This end-to-end discovery and follow-up campaign bodes well for future searches in this post-detection era of gravitational waves.
\end{abstract}


\section{Introduction}
The direct detection of gravitational waves (GW) marks the dawn of a new era \citep{GW150914-DETECTION}. It is widely agreed that the detection and study of the anticipated electromagnetic (EM) counterparts will vastly enrich the science returns for the field of GW astronomy. The photometric discovery of the EM counterpart will give a precise location and a spectrum of the host galaxy will give a precise redshift. This will enable a more accurate measurement of basic astrophysical properties such as the luminosity and energetics of this strong-field gravity event. If the spectrum is timely, it may also solve the long-standing mystery of the unknown sites of r-process nucleosynthesis.

The inherent challenge is that the two advanced GW interferometers, due to the low frequency of operation, give very poor on-sky localization \citep{TwoDetectorsLocalization1,TwoDetectorsLocalization2, TwoDetectorsLocalization3, TwoDetectorsLocalization4}. Nevertheless, the prospect of finding electromagnetic counterparts by searching large sky areas is promising as the search methodology is steadily improving --- from early efforts in the enhanced LIGO S6 run \citep{S6opticalEM}, to proof-of-concept localizations of coarse {\it Fermi} gamma-ray bursts \citep{iPTF13bxl,FermiPTF}, to a score of EM facilities promptly responding to GW150914 \citep{GW150914-EMFOLLOW}. 

At the time of the GW150914 trigger, there was no information disclosed on the nature of the event i.e. whether it was a binary black hole merger or binary neutron star merger or something else (GCN~\citegcn{18330}).  Many facilities undertook a search for an electromagnetic counterpart (e.g., \citealt{gbm,swift,pspessto,decam}). Months later, after offline analysis, the event was identified as a binary black hole merger (GCN~\citegcn{18858}).

Here, we present the intermediate Palomar Transient Factory (iPTF) follow-up effort. iPTF uses the Samuel Oschin 48-inch telescope on Palomar mountain equipped with the CFH12K camera with a field-of-view of 7.1 deg$^{2}$ \citep{lkd+09}.  Our motivation was to look for an optical counterpart powered by free neutron decay (\citealt{mbg+15}), or heavy element radioactive decay \citep{kfm15,mf14}. We describe the sky area coverage, candidate identification, spectroscopic classification and panchromatic follow-up.  We conclude with our plans for a way forward.  

\section{Identifying Candidates}

\begin{figure*}[!hbt] 
\centering
\includegraphics[width=0.8\textwidth]{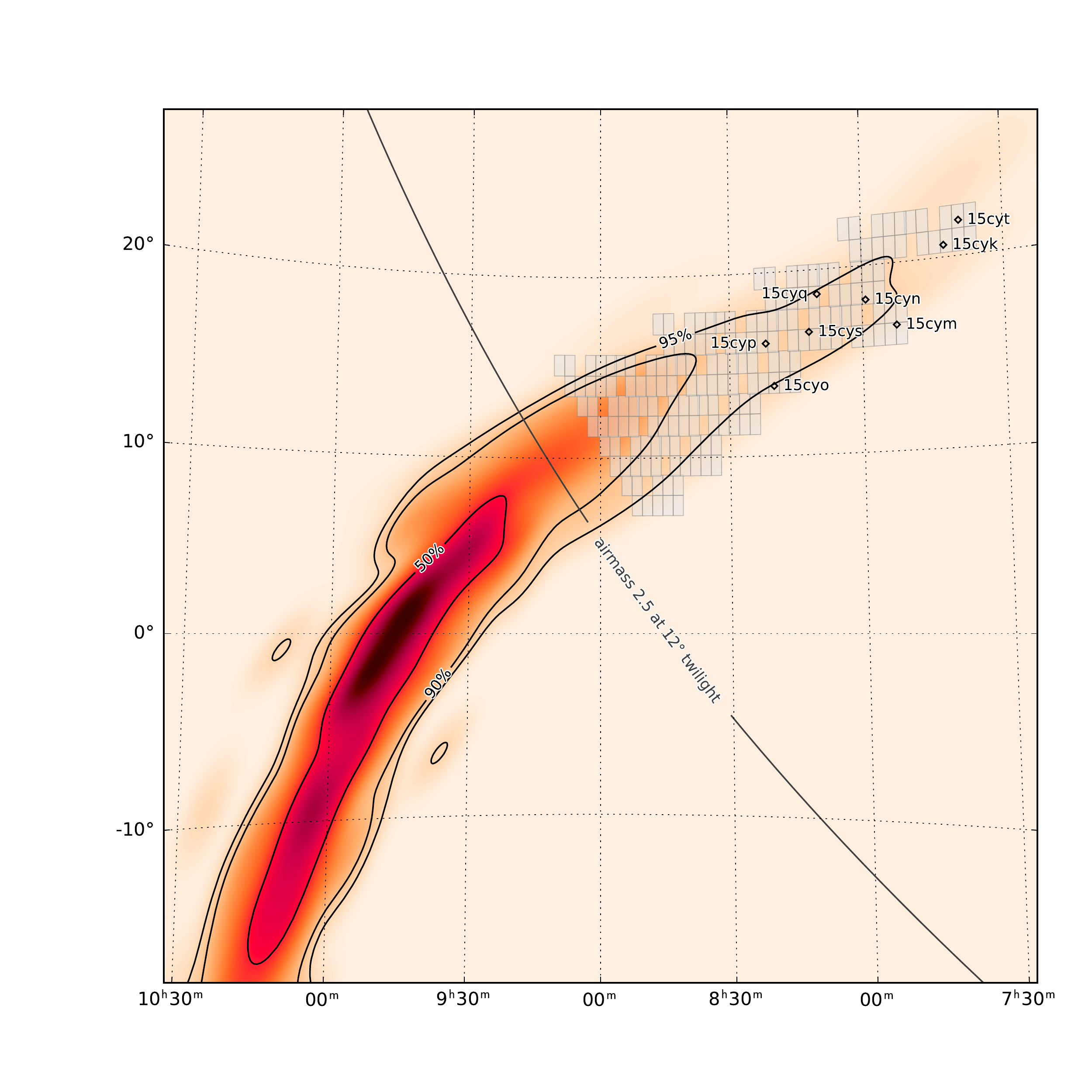}
\caption{\small iPTF coverage map (gray tiles) of GW150914. The color coding and contours denotes GW probability. Due to the sun-angle and elevation constraints, we were only able to image the westernmost region of the localization. Eight candidates were identified and classified.}
\label{fig:localization}
\end{figure*}

On UT 2015 September 16 03:17, the iPTF Target of Opportunity Marshal automatically responded to the gravitational wave trigger alert G184098 (later named GW150914). It immediately notified the team via phone calls and SMS alerts that there had been a GW trigger. It also computed that due to the sun angle constraint and elevation constraints, Palomar would only be able to access 2.5\% of the enclosed probability by tiling 126 deg$^{2}$ just before sunrise at high airmass (Figure~\ref{fig:localization}). This total area calculation takes into account the two non-working CCDs and the gaps between the CCDs.  The small containment probability was because the southern mode of the updated
(``LIB") localization was too far south to be observable from Palomar,
whereas most of the northern mode rose only after 12 degree morning twilight. Clouds did not co-operate and the Palomar 48-inch dome remained closed the first night after trigger. However, the next night (UT Sep 17), we imaged 18 fields covering this area with exposures of 1\,min (See details in Table~\ref{tab:obslog}; GCN~\citegcn{18337}). The scheduling and choice of tiles was further optimized applying the algorithm described in \citealt{iPTFScheduling}. A second epoch with a baseline separation of 30\,min ($\pm$ 1\,min) was obtained for 13 fields. 
 
Within minutes of obtaining the data, our automated real-time image subtraction pipeline started loading candidates into our database. We have two, independent real-time pipelines -- one running at the National Energy Research Scientific Computing Center (NERSC) using the HOTPANTS image subtraction algorithm \citep{nck15} and the other running at the Infrared Processing and Analysis Center (IPAC) using the PTFIDE algorithm \citep{masci16}. Due to the dynamic nature of the optical sky, the candidate list was dominated by false positive transients unrelated to the gravitational wave trigger.  A total of 127676 candidates were loaded into the NERSC database and 32576 in the IPAC database. Our automated machine-learning-aided  filtering algorithms rejected the moving objects in our solar system, variable stars in the Milky Way as well as subtraction artifacts. A list of 13 candidates were presented on a dynamic web portal for human vetting. 

We have been refining our software algorithms that quickly sift through the large number of candidates during our {\it Fermi} Gamma-ray Burst Monitor afterglow search effort \citep{FermiPTF}.  The EM-GW challenge has some similarities and some differences.  The similarities are that we need to continue to reject foreground asteroids/variable stars and background supernovae/active galactic nuclei. The differences are that compared to a Gamma Ray Burst afterglow, the EM-GW counterpart may be relatively fainter and/or slower and/or redder. Knowing that the EM counterpart is relatively nearby due to the advanced LIGO sensitivity helps further reduce false positives.  

The following are some rejection criteria:
\begin{enumerate}
\item Movement in detections in two epochs separated by at least 15\,min suggesting the candidate is an asteroid
\item Past history of eruption in PTF/iPTF data (baseline of six years) suggesting the candidate is an old transient
\item Previously known radio source or X-ray source suggesting the candidate is an active galactic nucleus
\item Previously known optical or infrared point source underneath the position suggesting the candidate is a stellar flare 
\end{enumerate}

The following criteria lead to flags for follow-up spectroscopy, additional photometry and/or multi-band follow-up:
\begin{enumerate}
\item Host galaxy (within 100\,kpc of transient) with spectroscopic redshift $<$0.05 (or photometric redshift $<$0.1) --- this is motivated by advanced LIGO's sensitivity limit to binary neutron star mergers 
\item Photometric evolution on hour timescale ($>$0.2\,mag) or day timescale ($>$\,0.5mag) or one-week timescale ($>$1\,mag) --- this serves as a strong discriminant against old supernovae. We note that this flag was not applied for GW150914 as all candidates of interest were spectroscopically classified within two hours. 
\item Hostless candidates with no counterpart in deep iPTF reference co-adds --- even though these are unlikely to be local, we flag these events as they are relatively rare.
\end{enumerate}

To quantify the relative efficacy of each criterion, we discuss the most severe cuts in order of severity by applying each criterion independently. Of the 127676 candidates in our NERSC pipeline, only 1007 candidates (0.8\% selection) are selected as being coincident with a galaxy within 200\,Mpc, hence this is the most severe cut. 5803 candidates (4.5\%) are selected as passing our machine learning cuts (we now have three generations of machine learning algorithms; see details in \citealt{umaa14,brink13}). 15624 candidates (12.2\%) are selected as having two detections separated by 30\.min in the same night. 78951 candidates (62\% selection) are selected as not having an optical point source in the reference image. Similarly, in our IPAC pipeline, we had a total of 32576 candidates. Of these, 24699 did not match a star (75.8\% selection),  5302 had two detections (16.2\% selection) and 1964 passed our machine learning cut (6.0\% selection).

In practice, these criterion are not all applied simultaneously and the candidates selected for human vetting are the result of a more complex database query. For example, prior to human vetting, we do not require coincidence with a nearby galaxy and we do not require any light curve properties.  For the five fields where a second epoch was not completed on the same night, we did a manual search requiring a local universe match, found 2 candidates that were both rejected as known asteroids. After human vetting of 13 candidates,  5 candidates were rejected as they showed past history of variability in the PTF data. 
In summary, our team flagged 8 candidates for further follow-up in our marshal database (see Table~\ref{tab:candidates}). 
Next, we describe the prompt follow-up that was undertaken to investigate whether any of the candidates was associated with GW150914 (GCN~\citegcn{18341}). 

\section{Spectroscopic Follow-up}
\begin{figure*}[!hbt] 
\centering
\includegraphics[width=\textwidth]{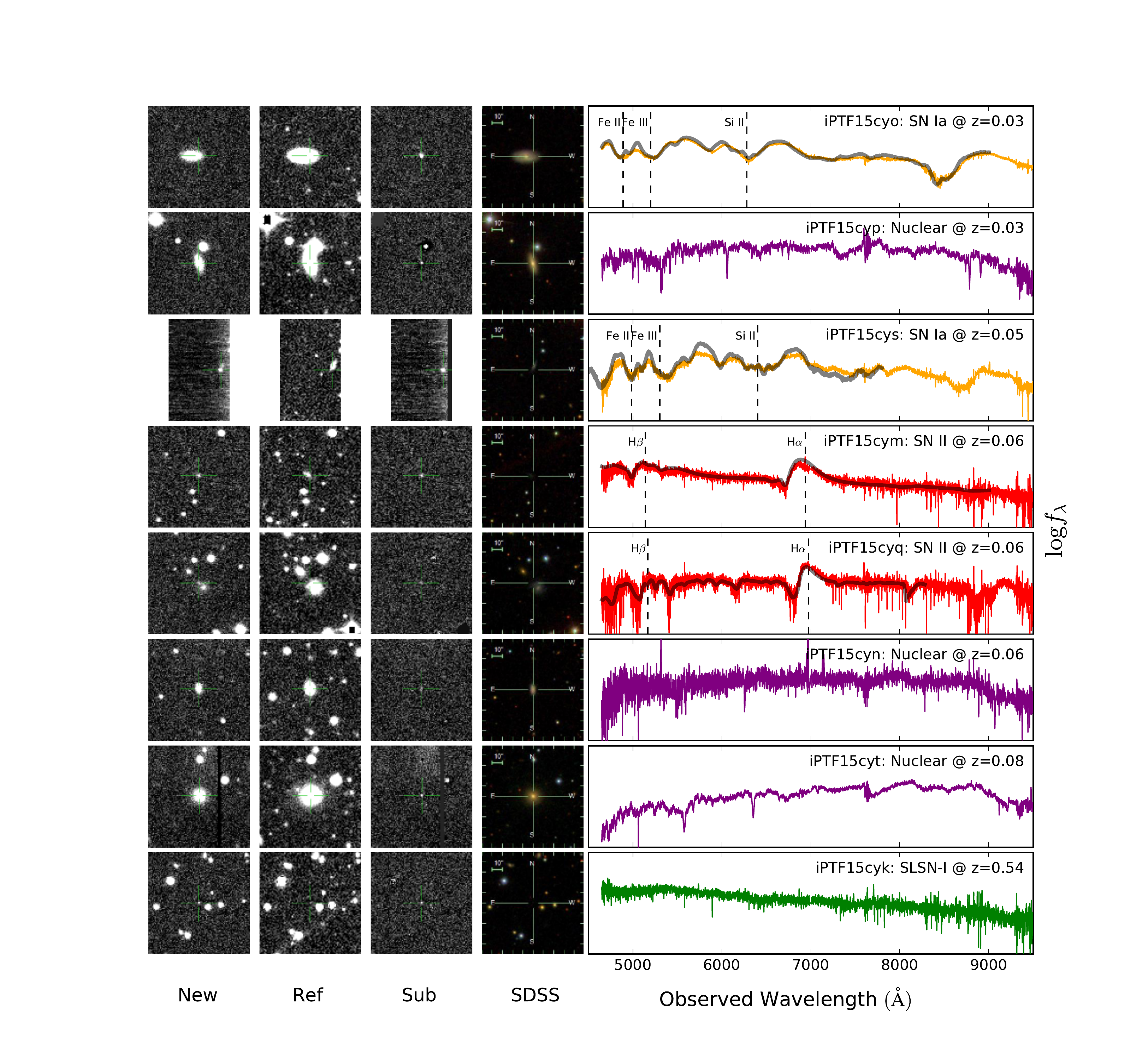}
\caption{\small Keck II/DEIMOS classification spectra of eight iPTF candidates obtained within 2 hours of discovery. Also shown, from left to right, the P48 discovery image, reference image, subtraction image and SDSS thumbnail around each candidate location. Colors denote spectroscopic class: SN Ia (red), SN II (blue), Nuclear (purple), SLSN I (green). Overplotted in gray lines is the best match from a supernova spectra library (SN1996X for iPTF15cyo, SN2004eo for iPTF15cys, SN1999M for iPTF15cym, SN2004et for IPTF15cyq). Additional follow-up data was needed to classify iPTF\,15cyk as a SLSN I (see Figure~\ref{fig:15cyk}).}
\label{fig:spec}
\end{figure*}

Since Hawaii is west of Palomar Observatory, sunrise was three hours later and we were able to obtain spectra of all eight candidates in less than 2 hours from discovery (Figure~\ref{fig:spec}). We emphasize that iPTF has routinely been obtaining spectroscopic classification on the same night as discovery, totaling 165 transients with spectra within 12 hours, thus far. We observed with the DEep Imaging Multi-Object Spectrograph  (DEIMOS; \citealt{deimos}) mounted on the Keck II telescope. We used the low resolution 600 ZD grating, giving spectral coverage between 4650\AA\, and 9600\AA\, with a resolution of 3.5\AA\, (full width at half maximum). Our spectra are shown in Figure~\ref{fig:spec}.  A priori, since we searched 126 deg$^{2}$ to a depth of 20.5\,mag, we expect $\approx$3.2 supernovae using the rates in \citealt{li11} (and assuming that supernovae are brighter than $-$17\,mag for 1\,month i.e. a volume out to z=0.075). 


We cross-matched our spectra with a library of supernovae spectra augmenting the \texttt{superfit} software \citep{superfit}.  Our classifications are in Table~\ref{tab:candidates}. We found two Type Ia supernovae (SN Ia), two hydrogen-rich core-collapse supernovae (SN II), three nuclear candidates (e.g. weak AGN where the spectrum is dominated by the host galaxy), and one hostless transient with initially unclear classification (iPTF15cyk).  Offline processing of the three nuclear candidates also shows past history of photometric variability in the PTF data, which is consistent with the AGN hypothesis. The spectrum of iPTF15cyk was dominated by a blue continuum, with narrow lines suggesting a redshift of 0.539 (which would imply a very luminous transient). Since the nature of the GW source was unclear, we decided to obtain additional spectroscopic and multi-wavelength follow-up. 



\section{Radio and X-ray Follow-up}

We observed iPTF15cyk and the necessary calibrators with the Karl G. Jansky Very Large Array (VLA; \citealt{perley09}) in its D and DnC configurations. The observations were performed in C-band ($\approx 6$\,GHz central frequency) under our Target of Opportunity program (VLA/15A-339; PI: Corsi). VLA data were reduced and imaged using the Common Astronomy Software Applications (CASA) package. In Table~\ref{tab:15cyk}, we report the $3\sigma$ upper-limits derived for iPTF15cyk using the full 2 GHz bandwidth (GCN~\citegcn{18914}). 

If the host galaxy redshift was confirmed, iPTF15cyk could be a super luminous supernova (SLSN) with absolute magnitude brighter than $-$22\,mag. Radio and X-ray emission from super luminous SNe may arise from interaction with the circumstellar medium (CSM; see e.g. \citealt{ofc+13}). In an alternate model, superluminous supernovae could be powered by the spin-down of a nascent magnetar inside the supernova ejecta \citep{kb10}, which may also produce X-ray emission \citep{mvb14}.

However, such emission is likely to be very sensitive to the exact properties of the CSM including density profile and homogeneity. In dense CSM environments, free-free absorption can suppress the radio emission at early times. Thus, chances for a detection are maximized by observing after maximum light \citep{ofc+13}. Hence, we observed iPTF15cyk thrice between 1\,month and 4\,months after discovery.

We also observed the location of iPTF15cyk with the \textit{Swift} satellite \citep{gcg+04} beginning at 18:12 UT on 2015 September 18 ($\Delta t = 4.3$\,d after the GW trigger).  We do not detect any emission with the on-board X-Ray Telescope (XRT; \citealt{bhn+05}) to a 3$\sigma$ limit of $< 3.2 \times 10^{-3}$\,ct\,s$^{-1}$.  Assuming a power-law spectrum with a photon index of $\Gamma = 2$, this corresponds to an upper limit on the unabsorbed flux (0.3--10.0\,keV) of $f_{X} < 1.3 \times 10^{-13}$\,erg\,cm$^{-2}$\,s$^{-1}$.

Simultaneously we obtained images of the field with the Ultra-Violet Optical Telescope (UVOT; \citealt{rkm+05}) on-board \textit{Swift} in the $V$, $B$, $U$, $UVW1$, $UVM2$, and $UVW2$ filters.  No emission is detected at the location of iPTF15cyk.  For a 3\farcs0 aperture we place the following magnitude limits (AB system) at this time: $V > 19.29$; $B > 19.81$; $U > 20.62$; $UVW1 > 21.61$; $UVM2 > 22.27$; and $UVW2 > 22.42$.  These limits were derived using the revised UV zero points and time-dependent sensitivity from \citet{blh+11}.

\begin{figure*}[!hbt] 
\centering
\includegraphics[width=0.9\textwidth]{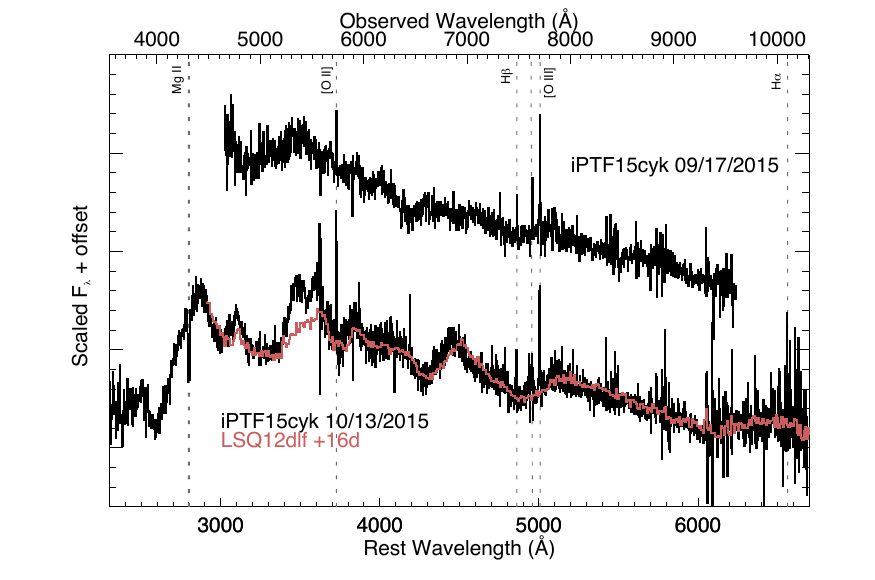}
\caption{ \small Spectral evolution of iPTF15cyk. The spectra show narrow lines from the host galaxy corresponding to z=0.539. The second spectrum matches a hydrogen-poor super luminous supernova, LSQ12dlf at +16d.
}
\label{fig:15cyk}
\end{figure*}

Additional spectroscopic follow-up of iPTF15cyk showed that it was a hydrogen-poor super luminous supernova (SLSN I; \citealt{quimby}) at z=0.539 (Figure~\ref{fig:15cyk}), similar to LSQ12dlf at +16\,d \citep{lsq12dlf}. The radio and X-ray upper limits were consistent with this classification. Given the high redshift, we concluded that this event was unrelated to GW150914. We note that the odds of finding a super luminous supernova were lower than the odds of finding other core-collapse or thermonuclear supernovae.  The snapshot rate is only $\sim$0.2 using the volumetric rate in \citealt{quimby13} (and assuming that SLSN are brighter than $-$21\,mag for 1\,month). Moreover, we have a total of only 6 events with z$>$0.5 (out of 2650 spectroscopically classified supernovae) in the six years of operating PTF/iPTF. 

\section{A Way Forward}
\begin{figure*}[!hbt] 
\centering
\includegraphics[width=0.7\textwidth]{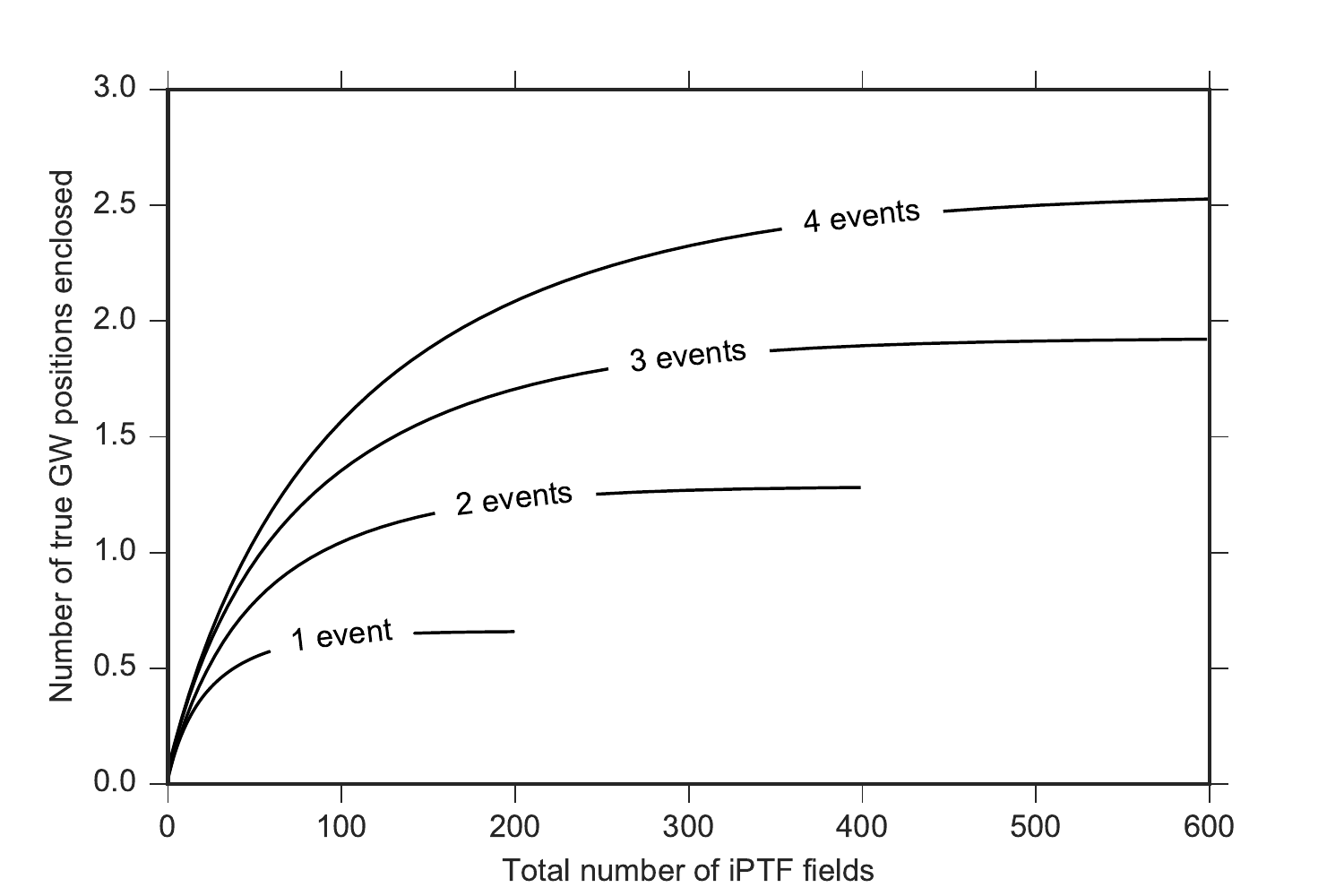}
\caption{ \small A simulation to compute the average number of times the true GW position would be enclosed in the iPTF imaged area as a function of the total number of iPTF fields imaged. Each iPTF field is 7.1 deg$^{2}$  and two 60\,s images of 150 fields can be obtained in a night. Thus, we  need to follow-up 2 GW events to have at least 1 in our imaged area. 
}
\label{fig:sim}
\end{figure*}

The post-detection era promises to be one of routine GW detections of binary neutron star mergers. 
With routine detections, the joint probability of $\sim\frac{1}{3}$ that the sun ($\sim\frac{2}{3}$), clouds ($\sim\frac{2}{3}$), and 
latitude ($\sim\frac{3}{4}$) simultaneously co-operate to identify the optical counterpart is not discouraging. 
Furthermore, given the location of Palomar Observatory in Southern California, relative to the location of the advanced LIGO interferometers, the time lag to respond is inherently less than an hour as we do not need to wait for the earth to rotate \citep{TwoDetectorsLocalization1}. Most of the GW150914 localization was not accessible from the Northern night sky. But, based on our simulations  \citep{TwoDetectorsLocalization2}, iPTF would include the true position of the GW source for an average of $\approx$1 out of  2 events assuming a total of 100 iPTF observations (see Figure~\ref{fig:sim}; each observation is two 60\,s exposures of 7.1 deg$^{2}$). 

As advanced LIGO ramps up in GW sensitivity, we are undertaking both hardware and software upgrades to improve EM sensitivity. In 2017, we plan to commission the Zwicky Transient Facility (ZTF{\footnote{http://ptf.caltech.edu/ztf}}; \citealt{k12,b14}), a 47\,deg$^{2}$ camera on the Palomar 48-inch, with a twelve times higher volumetric survey speed than iPTF. This increase in survey speed enables a faster cadence and deeper search for the optical counterpart (e.g., 22\,mag in 10\,min). The larger field-of-view may also be more robust to a shifting localization (e.g. for GW150914, our enclosed probability went from 2.5\% in the initial map to 0.2\% in the final map; see \citealt{GW150914-EMFOLLOW}). We are continuing to improve our software algorithms, e.g., better candidate filtering, image co-addition and more optimal image subtraction \citep{zog16}. We are continuing to complete our census of the local universe (CLU; Cook et al. in prep) as this 200\,Mpc galaxy catalog serves as the most severe filter for false positives (see examples in \citealt{nkg13}). 

\begin{figure*}[!hbt] 
\centering
\includegraphics[width=0.8\textwidth]{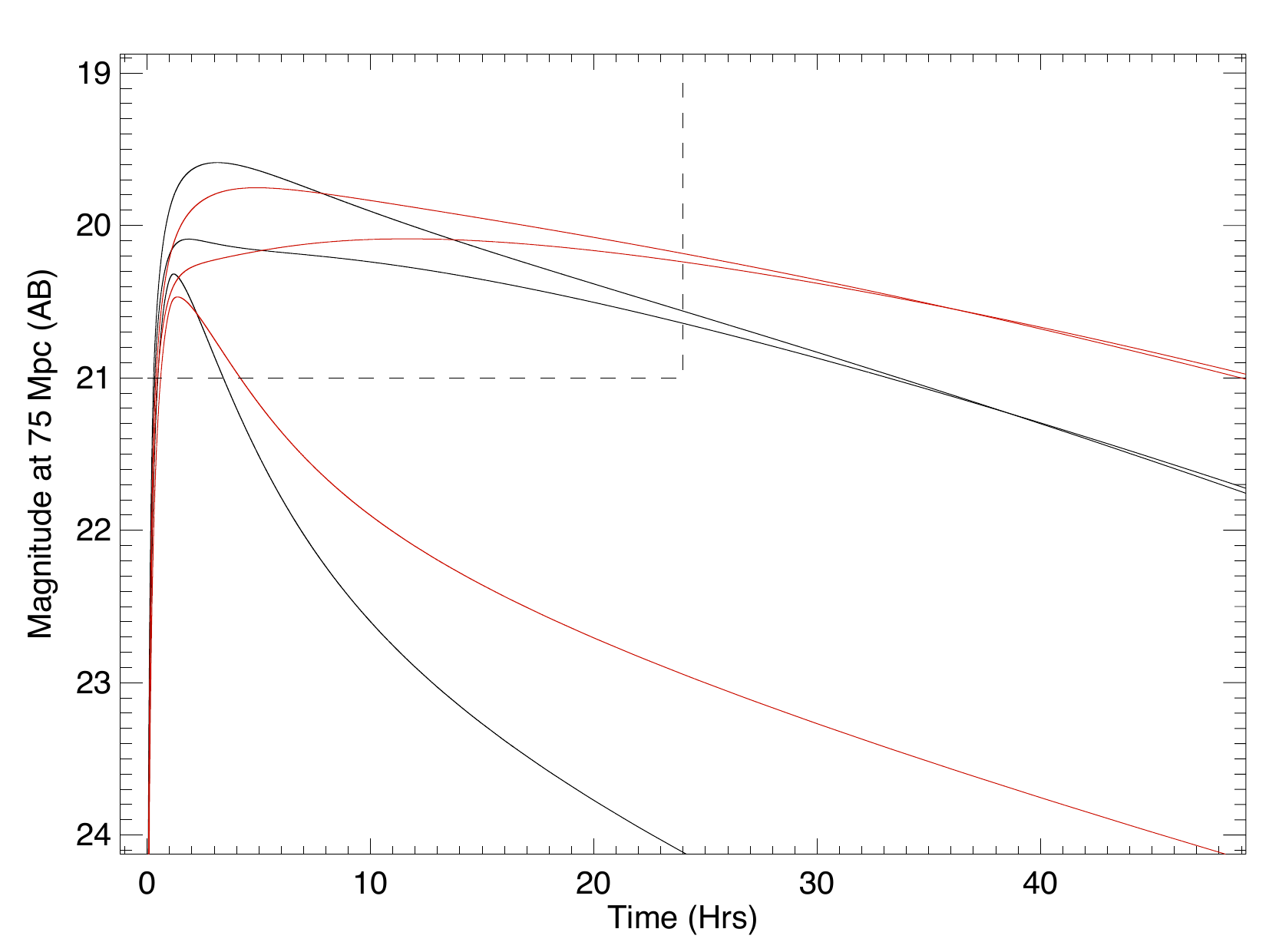}
\caption{\small Predicted optical counterpart based on free neutron decay \citep{mbg+15}. Black lines are g-band and red lines are r-band light curves at 75 Mpc (sensitivity limit of advanced LIGO to binary neutron star mergers in O1) . The three curves assume three different values for opacity and neutron mass to represent the fast, intermediate and slow light curve evolution cases i.e. ($\kappa_{r}$ = 30 cm$^{2}$gm$^{-1}$, M$_n$ = 3$\times$10$^{-5}$\.M$_{\odot}$), ($\kappa_{r}$ = 3 cm$^{2}$gm$^{-1}$, M$_n$ = 3$\times$10$^{-5}$\.M$_{\odot}$), ($\kappa_{r}$ = 3 cm$^{2}$gm$^{-1}$, M$_n$ = 3$\times$10$^{-4}$\.M$_{\odot}$). Note that g-band is more luminous at peak but decays faster. Horizontal dashed line denotes the sensitivity of iPTF in 60\,s. Vertical dashed line denotes the timescale within which follow-up is undertaken by the GROWTH program. 
}
\label{fig:neutrons}
\end{figure*}

Among the various models for electromagnetic emission from binary neutron star mergers, free neutron decay gives the most luminous optical counterpart (Figure~\ref{fig:neutrons}; \citealt{mbg+15}). Varying free neutron mass and opacity suggests that this counterpart may fade quickly, as much as 4\,mag in 24\,hours.  Thus, we are also systematizing our follow-up with the Global Relay of Observatories Watching Transients Happen (GROWTH{\footnote{http://growth.caltech.edu}}) program. The combination of a longitudinally distributed network of telescopes as well as multi-wavelength follow-up (VLA and \textit{Swift}) should effectively filter candidates on a 24 hour timescale. Obtaining a timely light curve, spectra and spectral energy distribution will unravel both the astrophysics and the astrochemistry of the EM counterpart. With this first gravitational wave detection, the 21st century gold rush \citep{k13} has officially begun!

\bigskip

Based on observations obtained with the Samuel Oschin Telescope 48-inch and the 60-inch Telescope at the Palomar Observatory as part of the intermediate Palomar Transient Factory (iPTF) project, a scientific collaboration among the California Institute of Technology, Los Alamos National Laboratory, the University of Wisconsin, Milwaukee, the Oskar Klein Center, the Weizmann Institute of Science, the TANGO Program of the University System of Taiwan, and the Kavli Institute for the Physics and Mathematics of the Universe. MMK, RL and YC acknowledge support from the National Science Foundation PIRE program grant 1545949. AAM acknowledges support from the Hubble Fellowship HST-HF-51325.01. PEN and YC acknowledge support from the DOE under grant DE-AC02-05CH11231, Analytical Modeling for Extreme-Scale Computing Environments. The National Radio Astronomy Observatory is a facility of the National Science Foundation operated under cooperative agreement by Associated Universities, Inc. AC and NP acknowledge support from NSF CAREER award 1455090. Part of the research was carried out at the Jet Propulsion Laboratory, California Institute of Technology, under a contract with NASA. MMK thanks Brian Metzger for providing us theoretical light curves for neutron-powered precursors. We thank the referee for constructive feedback. 

\begin{deluxetable*}{lccccccc}
\tablewidth{16cm}
\tablecaption{Candidates Flagged for Follow-Up}
\tablehead{
\colhead{Name} &
\colhead{RA (J2000)} &
\colhead{DEC (J000)} &
\colhead{Discovery time} &
\colhead{Mag (R-band)} &
\colhead{Minutes to Spectrum} &
\colhead{Classification} &
\colhead{Redshift}  
}

\startdata
iPTF15cyo & 8h 19m 56.18s  & +13d 52' 42.0" & 2015-09-17 05:54:55.6 & 17.75 $\pm$ 0.01 & 71   & SN Ia  (SN1996X-like, +23d)  & 0.029 \\ 
iPTF15cyp & 8h 21m 43.68s  & +16d 12' 42.0" & 2015-09-17 05:56:31.6 & 19.48 $\pm$ 0.05 & 125  & Nuclear & 0.028 \\ 
iPTF15cys & 8h 11m 55.59s  & +16d 43' 10.1" & 2015-09-17 06:05:16.6 & 17.84 $\pm$ 0.03 & 46   & SN Ia (SN2004eo-like, +22d) & 0.05  \\ 
iPTF15cym & 7h 52m 35.67s  & +16d 45' 59.6" & 2015-09-17 05:46:17.1 & 19.88 $\pm$ 0.20 & 113  & SN II (SN1999M-like, +5d)  & 0.055 \\ 
iPTF15cyq & 8h 10m 00.86s  & +18d 42' 18.1" & 2015-09-17 05:57:16.3 & 20.05 $\pm$ 0.10 & 39   & SN II  (SN2004et-like, +47d)  & 0.063 \\  
iPTF15cyn & 7h 59m 14.93s  & +18d 12' 54.9" & 2015-09-17 05:47:20.5 & 20.34 $\pm$ 0.28 & 124  & Nuclear & 0.062 \\ 
iPTF15cyt & 7h 38m 59.35s  & +21d 45' 43.2" & 2015-09-17 06:08:09.3 & 19.65 $\pm$ 0.09 & 82   & Nuclear & 0.078 \\ 
iPTF15cyk & 7h 42m 14.87s  & +20d 36' 43.4" & 2015-09-17 05:38:38.3 & 20.28 $\pm$ 0.12 & 97   & SLSN I (LSQ12dlf-like, +16d) & 0.539 \\ 
\enddata
\label{tab:candidates}
\end{deluxetable*}

\begin{deluxetable*}{llcc}
\tablewidth{16cm}
\tablecaption{Panchromatic follow-up of super luminous supernova iPTF15cyk}
\tablehead{
\colhead{Facility} &
\colhead{Epoch} &
\colhead{Frequency} &
\colhead{Flux Limit}  \\
\colhead{} & \colhead{UTC} & \colhead{} & \colhead{erg\,cm$^{-2}$\,s$^{-1}$ Hz$^{-1}$} 
}

\startdata
Swift/XRT & 18 Sep 2015 18:12   & 2\,keV & $<\,4.5 \times 10^{-32}$ \\
VLA & 15 Oct 2015 11:20:32-12:05:27  & 5.43 GHz  &  $<$\,2.6  $\times 10^{-28}$ \\
VLA & 06 Dec 2015 04:52:11-05:00:07 & 5.43 GHz & $<$\,2.3 $\times 10^{-28}$  \\
VLA & 20 Jan 2016  01:55:31-02:59:40 & 5.43 GHz & $<$\,2.3 $\times 10^{-28}$  \\
\enddata
\label{tab:15cyk}
\end{deluxetable*}

\begin{deluxetable*}{ccccccc}
\tablewidth{16cm}
\tablecaption{Observations Log}
\tablehead{
\colhead{PTF Field ID} &
\colhead{Central RA (J2000)} &
\colhead{Central DEC (J2000)} &
\colhead{Observation Time (UTC)} &
\colhead{Filter} &
\colhead{Airmass} &
\colhead{Limiting mag (5$\sigma$)} 
}
\startdata
     3050 &       132 &  7.875 & 2015-09-17 12:13:20.041 & R      & 2.7 & 20.4 \\  
     3050 &       132 &  7.875 & 2015-09-24 12:02:27.041 & R      & 2.3 & 20.3 \\  
     3050 &       132 &  7.875 & 2015-09-24 12:13:29.89  & R      & 2.1 & 20.3 \\  
     3154 & 129.80769 & 10.125 & 2015-09-17 12:11:38.141 & R      & 2.4 & 20.4 \\  
     3155 & 133.26923 & 10.125 & 2015-09-17 12:15:01.341 & R      &  2. & 20.2 \\  
     3257 & 127.57282 & 12.375 & 2015-09-17 12:03:09.29  & R      & 2.3 & 20.6 \\  
     3257 & 127.57282 & 12.375 & 2015-09-17 12:33:41.591 & R      &  1. & 19.6 \\  
     3258 & 131.06796 & 12.375 & 2015-09-17 12:09:56.041 & R      & 2.4 & 20.4 \\  
     3258 & 131.06796 & 12.375 & 2015-09-17 12:40:29.29  & R      & 1.9 & 18.2 \\  
     3259 & 134.56311 & 12.375 & 2015-09-17 12:16:43.241 & R      & 2.6 & 20.1 \\  
     3359 & 125.29412 & 14.625 & 2015-09-17 12:01:27.441 & R      & 2.1 & 20.6 \\  
     3359 & 125.29412 & 14.625 & 2015-09-17 12:32:00.191 & R      & 1.7 & 20.0 \\  
     3360 & 128.82353 & 14.625 & 2015-09-17 12:04:51.39  & R      & 2.2 & 20.5 \\  
     3360 & 128.82353 & 14.625 & 2015-09-17 12:35:23.591 & R      & 1.8 & 19.3 \\  
     3361 & 132.35294 & 14.625 & 2015-09-17 12:08:14.941 & R      & 2.4 & 20.4 \\  
     3361 & 132.35294 & 14.625 & 2015-09-17 12:38:46.99  & R      & 1.9 & 18.6 \\  
     3362 & 135.88235 & 14.625 & 2015-09-17 12:18:25.191 & R      & 2.5 & 20.2 \\  
     3459 & 119.40594 & 16.875 & 2015-09-17 11:52:55.541 & R      & 1.8 & 20.8 \\  
     3459 & 119.40594 & 16.875 & 2015-09-17 12:23:31.44  & R      & 1.5 & 20.7 \\  
     3459 & 119.40594 & 16.875 & 2015-12-03 09:36:36.833 & g      & 1.0 & 20.8 \\  
     3459 & 119.40594 & 16.875 & 2015-12-03 10:25:33.583 & g      & 1.0 & 20.8 \\  
     3459 & 119.40594 & 16.875 & 2015-12-09 08:44:28.483 & g      & 1.1 & 19.6 \\  
     3459 & 119.40594 & 16.875 & 2015-12-09 09:25:19.082 & g      & 1.0 & 20.7 \\  
     3459 & 119.40594 & 16.875 & 2015-12-15 09:16:17.383 & g      & 1.0 & 21.1 \\  
     3460 &  122.9703 & 16.875 & 2015-09-17 11:56:20.941 & R      &  1. & 20.6 \\  
     3460 &  122.9703 & 16.875 & 2015-09-17 12:26:54.891 & R      & 1.6 & 20.4 \\  
     3461 & 126.53465 & 16.875 & 2015-09-17 11:59:45.941 & R      &  2. & 20.6 \\  
     3461 & 126.53465 & 16.875 & 2015-09-17 12:30:18.641 & R      & 1.7 & 20.1 \\  
     3461 & 126.53465 & 16.875 & 2015-09-18 11:39:19.14  & R      & 2.4 & 20.6 \\  
     3461 & 126.53465 & 16.875 & 2015-09-18 12:00:28.941 & R      & 2.0 & 20.7 \\  
     3462 & 130.09901 & 16.875 & 2015-09-17 12:06:33.19  & R      & 2.2 & 20.6 \\  
     3462 & 130.09901 & 16.875 & 2015-09-17 12:37:05.491 & R      & 1.8 & 19.0 \\  
     3560 &     120.6 & 19.125 & 2015-09-17 11:54:37.991 & R      & 1.8 & 20.8 \\  
     3560 &     120.6 & 19.125 & 2015-09-17 12:25:12.741 & R      & 1.5 & 20.6 \\  
     3560 &     120.6 & 19.125 & 2015-09-18 11:13:55.441 & R      & 2.3 & 20.6 \\  
     3560 &     120.6 & 19.125 & 2015-09-18 11:37:37.341 & R      &   1 & 20.7 \\  
     3560 &     120.6 & 19.125 & 2015-12-03 09:38:18.533 & g      & 1.0 & 20.8 \\  
     3560 &     120.6 & 19.125 & 2015-12-03 10:27:16.483 & g      & 1.0 & 20.9 \\  
     3560 &     120.6 & 19.125 & 2015-12-09 08:46:08.433 & g      & 1.1 & 19.4 \\  
     3560 &     120.6 & 19.125 & 2015-12-09 09:26:59.082 & g      & 1.0 & 20.5 \\  
     3560 &     120.6 & 19.125 & 2015-12-15 09:17:59.533 & g      & 1.0 & 21.1 \\  
     3561 &     124.2 & 19.125 & 2015-09-17 11:58:03.64  & R      & 1.9 & 20.7 \\  
     3561 &     124.2 & 19.125 & 2015-09-17 12:28:36.291 & R      & 1.6 & 20.4 \\  
     3561 &     124.2 & 19.125 & 2015-12-03 09:46:51.133 & g      & 1.0 & 20.8 \\  
     3561 &     124.2 & 19.125 & 2015-12-03 10:35:47.883 & g      & 1.0 & 20.8 \\  
     3561 &     124.2 & 19.125 & 2015-12-09 09:35:19.683 & g      & 1.0 & 21.0 \\  
     3561 &     124.2 & 19.125 & 2015-12-15 09:26:36.183 & g      &  1. & 21.1 \\  
     3658 & 115.71429 & 21.375 & 2015-09-17 11:49:31.191 & R      & 1.6 & 20.9 \\  
     3658 & 115.71429 & 21.375 & 2015-09-17 12:20:07.791 & R      & 1.4 & 20.9 \\  
     3658 & 115.71429 & 21.375 & 2015-09-18 10:50:48.89  & R      & 2.3 & 20.6 \\  
     3658 & 115.71429 & 21.375 & 2015-09-18 11:12:13.34  & R      & 2.0 & 20.7 \\  
     3658 & 115.71429 & 21.375 & 2015-09-19 11:44:38.64  & R      & 1.6 & 20.7 \\  
     3658 & 115.71429 & 21.375 & 2015-09-19 12:26:26.091 & R      & 1.3 & 20.8 \\  
     3658 & 115.71429 & 21.375 & 2015-12-03 09:26:08.832 & g      & 1.0 & 20.8\\  
     3658 & 115.71429 & 21.375 & 2015-12-03 10:14:56.283 & g      & 1.0 & 20.9 \\  
     3658 & 115.71429 & 21.375 & 2015-12-09 08:41:08.233 & g      & 1.1 & 19.3 \\  
     3658 & 115.71429 & 21.375 & 2015-12-09 09:21:58.483 & g      & 1.0 & 20.7 \\  
     3658 & 115.71429 & 21.375 & 2015-12-15 09:12:53.733 & g      & 1.0 & 21.2 \\  
     3659 & 119.38776 & 21.375 & 2015-09-17 11:51:13.24  & R      & 1.7 & 20.9 \\  
     3659 & 119.38776 & 21.375 & 2015-09-17 12:21:49.841 & R      & 1.5 & 20.8  \\  
\enddata
\label{tab:obslog}
\end{deluxetable*}


\end{document}